\documentclass{article}
\usepackage{spconf,amsmath,graphicx}
\usepackage{stfloats}
\usepackage{amsfonts,amssymb,multirow}
\usepackage[numbers,sort&compress]{natbib}
\usepackage{enumitem}

\title{McNet: Fuse Multiple Cues for Multichannel Speech Enhancement}

\name{Yujie Yang$^{1, 2, \#}$,    Changsheng Quan$^{1, 2, \#}$,     Xiaofei Li $^{2, *}$ \thanks{$^{\#}$ equal contribution, * corresponding author}}
\address{$^{1}$ Zhejiang University, Hangzhou, China \\
$^{2}$ Westlake University \& Westlake Institute for Advanced Study, Hangzhou, China}

\begin{document}
%
\maketitle
\begin{abstract}
In multichannel speech enhancement, both spectral and spatial information are vital for discriminating between speech and noise. How to fully exploit these two types of information and their temporal dynamics remains an interesting research problem. As a solution to this problem, this paper proposes a multi-cue fusion network named McNet, which cascades four modules to respectively exploit the full-band spatial, narrow-band spatial, sub-band spectral, and full-band spectral information. 
Experiments show that each module in the proposed network has its unique contribution and, as a whole, notably outperforms other state-of-the-art methods.
\end{abstract}
\section{Introduction}
\label{sec:intro}
Speech enhancement aims to separate target speech from background noise, which is widely used in various applications, such as mobile telecommunication and hearing aids, and also serves as a front-end module for automatic speech recognition. Deep learning has been successfully used for single-channel speech enhancement \cite{8369155,9414177,9054254,hu20g_interspeech,LI2022108499}. These methods mainly exploit the differences in spectral patterns between speech and noise, and enhance noisy speech by building up the mapping network from noisy spectrogram to clean speech spectrogram. Among these methods, our previously proposed FullSubNet \cite{9414177} fuses a full-band and sub-band network, where the former learns the full-band spectral pattern (across-frequency dependencies) and the latter discriminates speech and noise based on the local (sub-band) spectral pattern and signal stationarity.  

Multichannel speech enhancement can further leverage spatial information. Beamforming (or spatial filtering) \cite{doi:https://doi.org/10.1002/9781119279860.ch10} is one leading technique, which applies a linear spatial filter to the noisy multichannel signals to suppress noise. The foundation of beamforming is that speech and noise have different spatial correlations. Beamforming is normally performed in narrow-band, as the spatial correlation of speech and noise can be formulated for each individual frequency. 
Neural beamforming \cite{7404829,erdogan16_interspeech} first uses a neural network to predict the speech presence of each time-frequency (T-F) bin, based on which the beamforming parameters, i.e., the steering vector of speech and the spatial covariance matrix of noise, can be estimated. 
To leverage the spatial information, \cite{7886357,8540037} integrate the inter-channel features, e.g., inter-channel phase/level differences (IPD/ILD), to the single-channel speech enhancement networks. Many works \cite{https://doi.org/10.48550/arxiv.1911.10791,9053989} directly process the multichannel signals to simultaneously exploit spectral and spatial information using some special network designs, such as the channel attention (CA) in \cite{9053989}. In our previous works \cite{li2019multichannel,https://doi.org/10.48550/arxiv.1911.10791}, a network is proposed to focus on the narrow-band spatial information, namely the difference of spatial correlation between speech and noise formulated in narrow-band.  \cite{https://doi.org/10.48550/arxiv.2206.13310} cascades a full-band network with the narrow-band network to exploit the full-band information simultaneously, and achieves the state-of-the-art (SOTA) performance.

Based on the researches discussed above, we have the following thoughts about speech enhancement. (1) Spectral information is important for discriminating between speech and noise, as speech and noise have different spectral patterns. Spectral information is present in both full-band and sub-band. Full-band spectral pattern is the major information adopted by most of the single-channel speech enhancement methods. Our previous works \cite{li20z_interspeech, 9414177, xiong22_interspeech} showed that sub-bands are also informative, in which the local spectral pattern and signal stationarity (normally speech is non-stationary and many types of noise are stationary) can be modeled. (2) Spatial information 
is also essential for discriminating between speech and noise, as normally speech is directional and spatially coherent while noise is diffuse or spatially less correlated. Spatial information is also present in both narrow-band and full-band. The spatial correlation of multichannel signals can be formulated in narrow-band. Spatial cues have extremely strong correlations across frequencies; for example, IPDs for different frequencies are all derived from the time delays of multichannel signals. (3) Spectral information and spatial information have their respective temporal dynamics. The temporal dependencies of spectral information reflect the signal content, while the temporal dynamic of spatial information reflects the spatial setting of source sources. Therefore, the amount of temporal contexts employed for spectral and spatial information may differ.

This work develops a multichannel speech enhancement network to fully exploit the spectral and spatial information mentioned above. The basic strategy is to exploit each type of information with a dedicated network. The single-channel FullSubNet \cite{9414177} and multichannel FT-JNF network \cite{https://doi.org/10.48550/arxiv.2206.13310} have proved that this strategy is effective for fusing different types of information. Specifically, the proposed multi-cue fusion network (named McNet) cascades four modules, including a full-band spatial, narrow-band spatial, sub-band spectral and full-band spectral module. Each module uses one layer of the LSTM network, by which the temporal dynamic of each type of information can be properly modeled. Compared with two SOTA networks, i.e., CA Dense U-net \cite{9053989} and FT-JNF  \cite{https://doi.org/10.48550/arxiv.2206.13310}, experiments show that the proposed network achieves notably better speech enhancement performance. 



\section{METHOD}
\label{sec:format}
 Multichannel speech signals can be written in the short-time Fourier transform (STFT) domain as:
\begin{align}
X_m(t,f)=S_m(t,f)+N_m(t,f), \label{2}
\end{align}
where $m\in[1,M]$, $f\in[0,F-1]$ and $t\in[1,T]$ denote the microphone, frequency and time frame indices, respectively, and $X_m(t,f)$, $S_m(t,f)$, and $N_m(t,f)$ are the STFT coefficients of noisy speech, clean speech and interference noise, respectively. Speech enhancement in this work aims to recover a single reference channel (denoted as $r$) of clean speech, i.e., $S_r(t,f)$, given the noisy multichannel signals.

\begin{figure}[t]
\centering
\includegraphics[width=1\columnwidth]{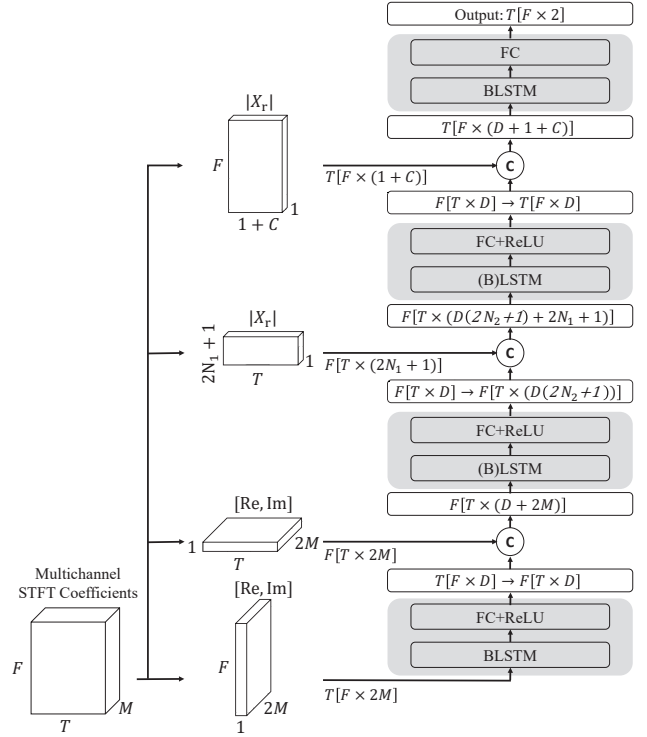}
\caption{\label{fig:network_structure}Diagram of McNet. The data dimensions  are represented in such a way: e.g., "$F[T\times2M]$" stands for $F$ independent sequences, each with time steps of $T$ and vector dimension of $2M$. The right arrow $\xrightarrow{}$ means reshape the dimensions. The letter $C$ within a circle means concatenation.}
\vspace{-.2cm}
\end{figure}

\subsection{Network architecture}
Fig. \ref{fig:network_structure} shows the diagram of the proposed network, which cascades four modules to exploit four different types of information, respectively. The information type is controlled by feeding different forms of the noisy signal to each module (together with the output of the previous module). Feeding the original noisy signal to each module also overcomes the problem that the previous modules may lose information. Each module is composed of a (B)LSTM and a linear layer. The first two modules exploit multichannel spatial information, while the last two exploit single-channel spectral information. 
Next, we will present the four modules one by one.


\subsubsection{Full-band spatial module}

This module learns the correlation of spatial cues, such as IPD and ILD, across frequencies. The input is the sequence of multi-channel STFT coefficients along the frequency axis at one frame ${\mathbf{X}}_{1}(t)=(\mathbf{x}(t,0),\cdots,\mathbf{x}(t,F-1))$, 
where
\begin{align}
     \mathbf{x}(t,f) = [&\text{Re}(X_{1}(t,f)),\text{Im}(X_{1}(t,f)),\cdots, \nonumber \\
     &\text{Re}(X_{M}(t,f)),\text{Im}(X_{M}(t,f))]^\mathrm{T} \in \mathbb{R}^{2M},
\end{align}
is the concatenation of multichannel STFT coefficients, $\text{Re}(\cdot)$ and $\text{Im}(\cdot)$ denote the real and imaginary parts of complex number, respectively, and $^{T}$ denotes vector transpose. The output of this module is denoted as $\mathbf{h}_{1}(t,f) \in \mathbb{R}^{D}$ for one T-F bin. 
Different frames are processed independently and share the same network. 

Running the LSTM recurrence along frequencies will make this module focus on the frequency dependencies in one frame. The single-frame spectral information also presents, but we think it is not very informative for speech enhancement, so this module focuses more on the full-band spatial information. This module does not learn any temporal dependencies, which is left for the next module. 


\subsubsection{Narrow-band spatial module}

The second module exploits the narrow-band spatial information and also builds up the temporal evolution of the output of module 1. 
The input is a temporal sequence at one frequency: ${\mathbf{X}}_{2}(f)=(\mathbf{x}_{2}(1,f),\cdots,\mathbf{x}_{2}(T,f))$, 
where 
\begin{align}
     \mathbf{x}_{2}(t,f) = [\mathbf{x}(t,f)^\mathrm{T}, \mathbf{h}_{1}(t,f)^\mathrm{T}]^\mathrm{T} \in \mathbb{R}^{2M+D}
\end{align}
is the concatenation of multichannel STFT coefficients and the output of module 1. The output of this module is denoted as $\mathbf{h}_{2}(t,f) \in \mathbb{R}^{D}$ for one T-F bin. Different frequencies are processed independently, and share the same network.   

The first two modules together follow a similar spirit as the FT-JNF network \cite{https://doi.org/10.48550/arxiv.2206.13310}. The major difference is that, besides the output of module 1, we also feed the narrow-band noisy signals to the second module, as the first module may lose some narrow-band information.

\subsubsection{Sub-band spectral module}

The third module exploits the sub-band (composed of multiple STFT frequencies) spectral information, mainly including the sub-band spectral pattern. The same as the second module, different frequencies are processed independently, and share the same network. 
For one frame, the input vector is composed of the spectral magnitude of the reference channel of one frequency and its $2N_1$ adjacent frequencies, and the output of the second module of this frequency and its $2N_2$ adjacent frequencies.
\begin{align}
\mathbf{x}_{3}(t,f) &= [\lvert X_{r}(t,f-N_{1}) \rvert,\cdots,\lvert X_{r}(t,f),\rvert,\cdots, \notag \\ &\lvert X_{r}(t,f+N_{1}) \rvert, \mathbf{h}_2(t,f-N_{2}) ,\cdots,  \mathbf{h}_2(t,f), \notag \\ & \cdots, \mathbf{h}_2(t,f+N_{2}) ] \in \mathbb{R}^{(2N_{1}+1)+D(2N_{2}+1)} ,
\end{align}
where $\lvert \cdot \rvert$ denotes absolute value. Its temporal sequence is ${\mathbf{X}}_{3}(f)=(\mathbf{x}_{3}(1,f),\cdots,\mathbf{x}_{3}(T,f))$. 
The output is denoted as $\mathbf{h}_{3}(t,f) \in \mathbb{R}^{D}$ for one T-F bin.

Signal stationarity is one important cue for discriminating between speech and noise, which is intensively leveraged in the spectral-subtraction-like methods \cite{5363084}. Signal stationarity is present in the single-channel narrow-band magnitude spectra. This sub-band module and the previous narrow-band module both involve the single-channel narrow-band magnitude spectra, and thus are both responsible for exploiting the signal stationarity.

\subsubsection{Full-band spectral module}

The last module exploits the full-band spectral pattern (and its temporal contexts) and collects/integrates the information obtained from the previous sub-band module across frequencies. The input sequence is organized along the frequency axis for each frame as ${\mathbf{X}}_{4}(t)=(\mathbf{x}_4(t,0),\cdots,\mathbf{x}_4(t,F-1))$.
The input vector for online speech enhancement is
\begin{align}
     \mathbf{x}_{4}(t,f) = [\lvert X_{r}(t-C,f) \rvert,\cdots,\lvert X_{r}&(t,f) \rvert,   
      \mathbf{h}_{3}(t,f)^\mathrm{T}]^\mathrm{T}  \notag\\ &\in \mathbb{R}^{C+1+D} 
\end{align}
where $C$ denotes the number of context frames. For the offline case, the future $C$ frames will also be concatenated into $\mathbf{x}_{4}(t,f)$, and the vector dimension will then be $2C+1+D$. The temporal dependencies of the full-band spectral pattern are mainly present for a small number of consecutive frames possibly within one phoneme, so this module concatenates several context frames and runs the LSTM recurrence along the frequency axis to learn better frequency dependencies.

Complex Ideal Ratio Mask (cIRM) \cite{7364200} is adopted as the learning target since it is well suitable for both single-channel and multichannel speech enhancement. Accordingly, the output of the last module is denoted as $\mathbf{y}_r(t,f) \in \mathbb{R}^{2}$ for one T-F bin, based on which the enhanced speech can be obtained.  





\subsection{Network configurations}
The proposed network can perform both online and offline speech enhancement. For the first and fourth modules that run the LSTM recurrence along the frequency axis, bidirectional LSTM is applied for both online and offline processing. For the second and third modules that run the LSTM recurrence along the time axis, unidirectional and bidirectional LSTMs are used for online and offline processing, respectively. 

The noisy multichannel signals are normalized before being processed by the network. Specifically, $X_m(t,f)$ is normalized with the magnitude mean $\mu$ of reference channel as $X_m(t,f)/\mu$. For offline processing, $\mu$ is simply computed as $\frac{1}{T}\frac{1}{F}\sum_{t=1}^{T}\sum_{f=0}^{F-1}\lvert X_{r}(t,f) \rvert$. For online processing, $\mu$ is recursively computed as $\mu(t)=\alpha \mu(t-1)+(1-\alpha)\frac{1}{F}\sum_{f=0}^{F-1}\lvert X_{r}(t,f) \rvert$,
where the smoothing parameter $\alpha = \frac{L-1}{L+1}$ is set to approximate a $L$-long smoothing window. 


\section{EXPERIMENTS}

\subsection{Experimental setup}
\textbf{Dataset}
Experiments are conducted on the simulated dataset of the third CHiME challenge \cite{7404837}. This dataset consists of 7,138, 1,640, and 1,320 utterances for training, development, and test, respectively. The sampling rate is 16 kHz. Data are recorded with a tablet device equipped with 6 microphones. Multichannel background noise are recorded in four environments, including a bus, cafe, pedestrian area, and street junction. 
We use the official script of the CHiME challenge to generate the simulated dataset. Except that, to make the simulated data more suitable for network training, the training data are generated on the fly with randomly selected noise clips with a signal-to-noise ratio (SNR) randomly chosen from the range of [-5,10] dB. Note that the multichannel speech signals for training and test are simulated by delaying single-channel speech utterances, thence they are reverberation free.  

\begin{table}[tbp]
\setlength\tabcolsep{4pt}
\renewcommand{\arraystretch}{1.3}
\caption{Performance of offline speech enhancement. \\ * means scores are quoted from the original papers.}
\label{tab:offline}
\centering
\resizebox{\linewidth}{!}{ 
\begin{tabular}{ccccc}
\hline
Method & NB-PESQ & WB-PESQ & STOI & SDR\\
\hline
Noisy & 1.82 & 1.27 & 87.0 
& 7.5\\
\hline
MNMF Beamforming * \cite{8673623} & - & - & 94.0 &16.2\\
Oracle MVDR & 2.49 & 1.94 & 97.0& 17.3\\
CA Dense U-net * \cite{9053989} & - & 2.44 & - &18.6\\
Narrow-band Net \cite{https://doi.org/10.48550/arxiv.1911.10791}& 2.74 & 2.13 & 95.0 & 16.6\\
FT-JNF \cite{https://doi.org/10.48550/arxiv.2206.13310}& 3.17 & 2.48 & 96.2 & 17.7\\
\hline
McNet (prop.) & \textbf{3.38} & \textbf{2.73} & \textbf{97.6} & \textbf{19.6}\\
\hline
\end{tabular}}
\vspace{-0.5cm}
\end{table}

\begin{table}[tbp]
\setlength\tabcolsep{4pt}
\renewcommand{\arraystretch}{1.3}
\caption{Performance of online speech enhancement. }
\label{tab:online}
\centering
\resizebox{\linewidth}{!}{ 
\begin{tabular}{ccccc}
\hline
Method & NB-PESQ & WB-PESQ & STOI & SDR\\
\hline
Noisy & 1.82 & 1.27 & 87.0 
& 7.5\\
\hline
Narrow-band Net \cite{https://doi.org/10.48550/arxiv.1911.10791}& 2.70 & 2.15 & 94.7 & 16.0\\
FT-JNF \cite{https://doi.org/10.48550/arxiv.2206.13310}& 2.80 & 2.23 & 95.4 & 16.9\\
\hline
McNet (prop.) & \textbf{3.29} & \textbf{2.67} & \textbf{97.2} & \textbf{19.0}\\
\hline
\end{tabular}}
\vspace{-0.4cm}
\end{table}

\textbf{Configurations}
STFT is performed using a 512-sample (32ms) Hanning window with a frame step of 256 samples. We set the length of utterances used for training fixed as $T=192$ frames (around 3s). The numbers of hidden units are set to 128, 256, 384, and 128 for (each direction of) the LSTM layer of four modules, respectively. The output dimensions of the first three modules are all set to $D=64$. The number of adjacent frequencies used for the third module are $N_1=3$ and $N_2=2$. The number of context frames for the fourth module is $C=5$. The length of the smoothing window is set to $L=192$ to be consistent with the length of training utterances. We take channel No. 5 as the reference channel.

Adam is used as the optimizer, and gradient clipping with maximum $L_{2}$-norm of 5 is applied. An exponential decay learning schedule is used with an initial learning rate of 0.001 and a decaying factor of 0.992. The batch size is set to 3. We train our models until convergence, which takes almost 500 epochs. 
Four evaluation metrics are used: NB-PESQ and WB-PESQ \cite{941023}, STOI \cite{5713237} and SDR \cite{1643671}. Code and some speech examples are available on our website \footnote{https://github.com/Audio-WestlakeU/McNet}.

\textbf{Comparison methods}
We compare with the following multichannel speech enhancement methods. 
(1) MNMF Beamforming \cite{8673623} is an unsupervised speech enhancement method, first using multichannel nonnegative matrix factorization (MNMF) to estimate the spatial covariance matrices of speech and noise, then performing beamforming for speech enhancement. 
(2) Oracle Minimum Variance Distortionless Response (MVDR) \footnote{https://github.com/Enny1991/beamformers} uses the ground-truth spatial covariance matrices of noise, which is supposed to obtain the upper-bound performance of all MVDR-based beamforming methods.
(3) CA Dense U-net \cite{9053989} uses Channel Attention (CA) to perform non-linear spatial filtering, in the framework of Dense U-Net. 
(4)  Narrow-band Net \cite{https://doi.org/10.48550/arxiv.1911.10791} is one of our previous works that uses two layers of LSTM to only exploit the narrow-band spatial information (like the second module of the proposed network).
(5) FT-JNF \cite{https://doi.org/10.48550/arxiv.2206.13310}: Kristina et al. revised the Narrow-band Net \cite{https://doi.org/10.48550/arxiv.1911.10791} by replacing the first LSTM with an along-frequency LSTM to further exploit the full-band information (like the first and second modules together of the proposed network).

\begin{table}[t]
\setlength\tabcolsep{4pt}
\renewcommand{\arraystretch}{1.3}
\caption{Results of Ablation Studies.}
\label{tab:ablation_studies}
\centering
\resizebox{\linewidth}{!}{ 
\begin{tabular}{lcccc}
\hline
Method & NB-PESQ & WB-PESQ & STOI & SDR\\
\hline
Noisy  & 1.82 & 1.27 & 87.0 
& 7.5\\
\hline
McNet (prop.) & {3.29} & {2.67} & {97.2} & {19.0}\\
- full-band spatial &3.24&2.61&97.1&18.7\\
- narrow-band spatial &3.16&2.51&96.7&18.3\\
- sub-band spectral &3.25&2.57&96.9&18.2\\
- full-band spectral &3.18 &2.52 &96.7&18.5\\
\hline
\end{tabular}}
\vspace{-0.4cm}
\end{table}

\subsection{Speech enhancement results}

Table~\ref{tab:offline} shows the offline speech enhancement results. CA Dense U-net, FT-JNF and the proposed model outperform MNMF beamforming and even oracle MVDR, which shows the superiority of supervised methods for multichannel speech enhancement. Compared to Narrow-band Net, FT-JNF largely improves the performance by using the along-frequency LSTM layer to exploit full-band  information. On top of FT-JNF, the proposed model adds two single-channel LSTM layers dedicated to respectively exploit the sub-band and full-band spectral information, which again largely improves the performance. This verifies that the spectral information is complementary to the spatial information, as long as the spectral information can be fully used. 
Table~\ref{tab:online} shows the online speech enhancement results. Not surprisingly, the online performance measures are not as good as the ones of the offline case. The proposed model still achieves excellent results, which are even better than the offline results of comparison methods. This means, when the information of multiple cues (spatial and spectral; full-band, sub-band and narrow-band) are fully exploited, speech enhancement can be well conducted even without using future information. 

\textbf{Ablation studies} 
Table \ref{tab:ablation_studies} shows the results of ablation studies for the online speech enhancement case. The effectiveness of the four modules is verified by removing each of them. It can be seen that the performance measures are decreased when any module is removed, which indicates that there is no module being redundant in the proposed network. Among the four modules, the narrow-band spatial and full-band spectral modules seem more important. 

\section{CONCLUSION}
In this paper, we propose a multi-cue fusion network named McNet, for multichannel speech enhancement. It cascades four modules dedicated to respectively exploit the full-band spatial, narrow-band spatial, sub-band spectral, and full-band spectral information. This cascading architecture can effectively accumulate information. The proposed network achieves excellent speech enhancement performance, especially for the online case. 

\vfill\pagebreak
\clearpage

\bibliographystyle{IEEEbib}
{\small
\bibliography{strings,refs}}

\end{document}